\documentstyle[12pt,epsf]{article}
\begin{document}
\hfill LPTHE Orsay 94/43\\
\vspace{1cm}
\begin{center}
DISORIENTED CHIRAL CONDENSATES
\footnote{Talk at XXIX
Rencontres de Moriond, M\'eribel,
March 1994}\\
\vspace{0.9cm}
Andr\'e KRZYWICKI \footnote{E-mail
: krz@qcd.th.u-psud.fr}\\
Laboratoire de Physique Th\'eorique
 et Hautes Energies, B\^{a}t. 211,\\
Universit\'e Paris-Sud, 91405 Orsay, France \footnote{Laboratoire associ\'e au
C.N.R.S.}\\
\vspace{1cm}
ABSTRACT\\
\end{center}
\vglue 0.1cm
{\rightskip=3pc
\leftskip=3pc
\noindent
The idea that a bubble of misaligned vacuum
 is sometimes produced in high energy collisions
 is reviewed.}

\vspace{6.5cm}
\pagebreak
This talk is devoted to the popular topic of
 disoriented chiral condensates. The idea has
 been put forward in refs. 1 - 3 and has
received a further boost from ref. 4. At present,
 several dozens of papers on this subject are
 circulating and the literature is constantly
growing. In the following, results from ref. 5
will often be used.
\par
As you all know, QCD has an approximate SU(2)
 $\times$ SU(2) global symmetry. This symmetry is spontaneously broken and the
relevant part of
the order parameter
 $\psi^a \bar{\psi}^b$ ($a,b = 1,2$) is a vector
 $\cal M =(\sigma, \vec{\pi})$ transforming under
 the O(4) subgroup of SU(2) $\times$ SU(2). The
physical vacuum is a medium where $\cal M$ points
 in the $\sigma$ direction, because the pion mass
 $m_\pi$ is non-vanishing. However, the pion mass
 is small and the energy cost of tilting $\cal M$
is also small, of the order of $m^2_\pi f^2_\pi /
 ({\rm unit \; volume})$, where $f_\pi$ is the
pion decay constant. The medium where
$\cal M$ is misaligned has been baptized by
Bjorken a "disoriented chiral condensate" (DCC).
 It is clear that DCC must be insulated from
the physical vacuum and that its existence can
 only be ephemeral.
$$
\epsfbox{fig1.ps}
$$
\begin{center}
Fig. 1 \\
\end{center}
\vspace{0.6cm}
\par
Fig. 1 provides an intuitive picture of what
 might happen in some high-energy hadronic and/or
 nuclear collisions. The energetic debris form
a shell shielding the inner region, where DCC
resides, from the outer physical vacuum. The
phenomenon is truly interesting only when the
pion field in the inner region is coherent,
i.e. essentially classical. It is then expected
 that it points in a random direction in isospace.
 This implies, in turn, a striking experimental
signature. Let $f$ denote the fraction of neutral
 pions among all pions resulting from the decay
of DCC. The random variable $f$ is distributed
according to the law

\begin{equation}
{{dP(f)} \over {df}} = {1 \over {2\sqrt{f}}} ,
\label{law}
\end{equation}

\noindent
written first explicitly in ref. 3 (but known
earlier to Bjorken). The original derivation of
 (\ref{law}) is semi-classical, but there exists
  $^{6)}$ a quantum mechanical version of the
argument, which makes it clear that there is
no conflict between (\ref{law}) and isospin
conservation. Thus, there is 10\% probability
that in a cluster of 100 pions only one is
neutral! This is very far from any naive
statistical expectation and is of immediate
 phenomenological interest: DCC might perhaps
 be an explanation for the mysterious Centauro
 events observed by cosmic ray people~$^{7)}$.
 Furthermore, if DCC is indeed produced in high
 energy collisions it may convey relatively clean
information about the initial dense state, since
 it is so well signed.
\par
We shall now discuss the dynamics of DCC in
more detail, using the classical equations
of motion of the linear $\sigma$ model (cf.
 ref. 4).  The Lagrangian is

\begin{equation}
L = {1 \over 2}[(\partial \sigma)^2 +
 (\partial \vec{\pi})^2] - {\lambda \over 4}
 (\sigma^2 + \vec{\pi}^2 -1)^2 + H \sigma ,
\label{lagr}
\end{equation}

\noindent
For the sake of the argument we set
 $\lambda \gg H$ (the fields are dimensionless
and an overall dimensionfull factor has been
omitted, for simplicity of writing - thus
 $\lambda$ and $H$ have the dimension of
[mass]$^2$ ).
\par
In order to proceed analytically, we adopt
an idealization due originally to Heisenberg
 $^{8)}$ : we assume that at time $t = 0$,
the whole energy of the collision is localized
within an infinitesimally thin slab with infinite
 transverse extent (instead of a pancake shaped
 region). The symmetry of the problem then
implies that the fields can only depend on
the proper time $\tau = \sqrt{t^2 - x^2}$~,
 where $x$ is the longitudinal coordinate.
Notice, that $f = f(\tau)$ implies
$\partial^2 f = \ddot{f} + \dot{f}/\tau$ .
The action of the d'Alambertian produces both an "acceleration" and a
"friction" term,
respectively. Of course, there is no true
energy dissipation. The energy in a covolume
 decreases because of the expansion of the system.
\par
We further assume random initial conditions
for the fields and their gradients, such that
 chiral symmetry is initially unbroken:
${\langle \cal M \rangle} = {\langle
\dot{\cal M} \rangle} =$ 0 at $\tau =
 \tau_0$. This initial randomness can
perhaps result from a rapid cooling
("quench") of the quark-gluon plasma, as
 proposed in ref. 4, but we do not insist on this interpretation.
\par
The equations of motion are

\begin{eqnarray}
\ddot{\vec{\pi}} + \dot{\vec{\pi}}/\tau
 & = & - \lambda (\sigma^2 + \vec{\pi}^2
 - 1 ) \vec{\pi} \\
\ddot{\sigma} + \dot{\sigma}/\tau  & = &
 - \lambda (\sigma^2 + \vec{\pi}^2 - 1 ) \sigma + H
\label{eqsmot}
\end{eqnarray}

\noindent
{}From these equations one easily finds that

\begin{equation}
\vec{\pi} \times \dot{\vec{\pi}} = \vec{a}/\tau
\label{cvc}
\end{equation}

\noindent
and

\begin{equation}
\vec{\pi} \dot{\sigma} - \sigma
\dot{\vec{\pi}} = \vec{b}/\tau +
{H \over \tau} \int^\tau \vec{\pi}\;
 \tau \; d\tau
\label{pcac}
\end{equation}

\noindent
The isovectors $\vec{a}$ and $\vec{b}$
are integration constants. Eq. (\ref{cvc})
 is a consequence of the conservation of
the isovector current, while eq. (\ref{pcac})
reflects the partial conservation of the
iso-axial-vector current. The constants
$a$ and $b$ measure the initial strength
of these currents. The component of
$\vec{\pi}$ along $\vec{a}$ necessarily
vanishes, $\pi_a = 0$.
\par
One can parameterize the fields with one
radial and two angular variables:

\begin{eqnarray}
\pi_b & = & r \sin{\theta}\\
\pi_c & = &  r \cos{\theta} \sin{\omega} \\
\sigma & = &  r \cos{\theta} \cos{\omega}
\label{fields}
\end{eqnarray}

\noindent
where $\vec{c} = \vec{a} \times \vec{b}$.
Assuming, for a moment, that $H = 0$ one
finds that the angle $\omega$ is constant
 : $\omega = \arctan{(a/b)}$.
\par
The discussion of the quality of the
approximation $H = 0$ requires more space.
 One finds~$^{5)}$ that this approximation
 is realistic provided $\tau \ll b/\sqrt{H}$
 and $a \ll b$. Actually, the last condition
 only insures the planarity of the motion
and is not truly essential. We assume that
 this condition is satisfied because this
simplifies the discussion considerably:
the component $\pi_c$ of the pion field
is then always much smaller than $\pi_b$
and can be neglected altogether.
\par
Thus, we consider a dynamical system
described by one radial coordinate $r$
and by one angular coordinate $\theta$.
Consider the radial motion first.
\par
{\em \underline{The radial motion}}. In
this case the presence of the term $\propto H$
 in the equations of motion is an inessential
 complication. Setting $H = 0$ one gets after
 some algebra the following differential equation
 for $r$ (we neglect $a$ compared to $b$):

\begin{equation}
\ddot{r} + \dot{r}/\tau = {{b^2} \over
 {\tau^2 r^3}} - \lambda r (r^2 - 1)
\label{req}
\end{equation}

\noindent
This can be regarded as an equation of
motion of a material point in a time
dependent potential well with friction.
 The mechanical energy of the point
satisfies the inequality

\begin{equation}
{d \over {d\tau}} [{1 \over 2} \dot{r}^2
+ {\lambda \over 4} (r^2 - 1)^2 + {{b^2}
 \over {2\tau^2r^2}}] = - {{b^2} \over
{\tau^3r^2}} - {{\dot{r}^2} \over \tau} < 0
\label{ineq1}
\end{equation}

\noindent
and, therefore, decreases monotonically
while $r \to 1$. Linearizing the force
with respect to $r-1$ one obtains a Bessel
 equation whose large time solution reads

\begin{equation}
r = 1 + {\rm const} \cos{(\tau\sqrt{2\lambda}
 + \eta)}/(\tau \sqrt{2\lambda})^{{1 \over 2}}
\label{solr}
\end{equation}

\noindent
Eq. (\ref{solr}) exhibits damped oscillations
 around the equilibrium position $r = 1$.
 These oscillations have frequency $\propto
 \sqrt{2\lambda}$, which should be regarded,
 in this context, as large. Averaging over a
few periods of the radial motion one finds
the expectation value $\langle r \rangle =
1$. This value is reached in a relatively
short time, when the time dependent part
of the potential ceases to be important,
i.e. for $\tau \; \sim \; b/\sqrt{2\lambda}$.
\par
{\em \underline{The angular motion}}. In
the approximation $H = 0$, the equation
governing the angular motion is

\begin{equation}
\dot{\theta} = {b \over {r^2 \tau}}
\label{eqth}
\end{equation}

\noindent
Since the angular motion is generically
slower than the radial one, we replace in
 (\ref{eqth}) $r$ by its expectation value
$\langle r \rangle = 1$:

\begin{equation}
\dot{\theta} = {b \over \tau}
\label{approxth}
\end{equation}

\noindent
The solution is $\theta \approx b
 \ln{(\tau/\tau_0)}$, identical to
 that found in ref. 3 using the
 non-linear $\sigma$ model.
\par
When $H > 0$, (\ref{approxth}) is replaced by

\begin{equation}
\ddot{\theta} + \dot{\theta}/\tau +
 H \sin{\theta} = 0
\label{thex}
\end{equation}

\noindent
which is the equation of a pendulum
with friction. Now, the time derivative
 of the corresponding mechanical energy
satisfies the inequality

\begin{equation}
{d \over {d\tau}} [{1 \over 2}
 \dot{\theta}^2 + H(1 - \cos{\theta})]
 = - {{\dot{\theta}^2} \over \tau} < 0
\label{then}
\end{equation}

\noindent
This energy decreases monotonically,
until it becomes of the order of $H$.
 At that time the circular motion of
the pendulum turns into an oscillatory
 one. One can show $^{5)}$ that the
change of regime occurs when $\tau
\approx b/\sqrt{H}$ and that at large
 time the solution to (\ref{thex}) is

\begin{equation}
\theta \sim \sqrt{b} \cos{(\tau \sqrt{H}
 + \delta)}/(\tau \sqrt{H})^{1 \over 2}
\label{thsol}
\end{equation}

\noindent
Notice, that at large enough time
$\pi_b \approx \theta$. Actually,
(\ref{thsol}) describes free propagation
 of a pion with mass $\sqrt{H}$. It
can easily be seen that (\ref{thex})
reduces to the Klein-Gordon equation
for $\theta \ll 1$.
\par
Fourier transforming (\ref{thsol}) and
 squaring one gets the pion spectrum (per
 unit transverse area). One finds a
rapidity plateau with height $\sim b$.
 Of course, the existence of the rapidity
plateau is an artifact of the boost-invariant
 initial conditions and should not be treated
 seriously. The important point is that the
energy released in the decay of the condensate is proportional to the initial
strength $b$ of the iso-axial-vector current.
\par
Assuming, as in ref. 4, that the initial
 fluctuations of the fields and of their
gradients are Gaussian with variances
 $\sigma^2_{f,g}$ respectively, one finds
 $^{5)}$ the probability distribution of $b$ :

\begin{equation}
{{dP(b)} \over {db}} = {A \over
 {2(\pi\sigma_f \sigma_g  \tau_0 )^2}}
 K_1 \left( {b \over {\sigma_f \sigma_g
 \tau_0}} \right) \;, \; \; \; \; \; \;
 0 < a < A \ll b
\label{prob}
\end{equation}

\noindent
where $K_1(z) \sim \sqrt{\pi/2z} e^{-z}$
 for $z \to \infty$, is the modified Bessel
 function.
\par
Thus, with Heisenberg's idealization the
 problem becomes $1+1$ dimensional and can
 be solved analytically. At fixed time $t$
one finds quite naturally an inner region
 in which DCC resides, insulated from
the outer physical vacuum by two Lorentz
 contracted regions where $\cal M$ and
$\dot{\cal M}$ fluctuate around zero.
The {\em proper} time evolution of the
inner region is that of a simple
dynamical system. We have identified several
stages in the evolution
of this system. First there
is a short phase in which
angular motion and radial motion
are strongly coupled. This phase lasts for
a time $\tau \sim b/\sqrt{2\lambda}$.
Next, the fictitious particle rotates slowly,
while oscillating rapidly about the
equilibrium radial position $r = 1$.
At some point
there is  a transition from
circular to oscillatory motion, which
actually corresponds  to free
propagation of final state pions. The
transition takes place when the pion mass
 can no longer be
neglected and it occurs typically when
$\tau \sim b/\sqrt{H}$. The pion field
has a random orientation in
isospace (when $a \ll b$ it oscillates
along the direction of $\vec{b}$) and
therefore the fraction
of neutral pions is distributed according
to (\ref{law}).
\par
The two time scales, $(2 \lambda)^{-{1
\over 2}}$ and $H^{-{1 \over 2}}$, when
estimated using the phenomenological
values of the pion and $\sigma$ masses,
differ only by a factor of 4. This is
presumably not large enough to insure
a clean separation of regimes, unless
$b$ is large enough (say of order 3 to 5,
 or so). The analytic discussion sketched
 above indicates that the formation of an
observable  DCC is likely to be a rather
 natural but rare phenomenon, since the
probability distribution of $b$ decreases
 exponentially. Hence, it is not interesting
 to average over $b$.
\par
Once the simplifying assumptions made in
the above discussion are relaxed, the
 problem becomes rapidly untractable
analytically and one has to use a computer.
 Numerical calculations have been carried
out by several authors. In a $1+1$ dimensional
 scenario, but without boost invariance, it
 is found $^{9)}$ that the rapidity interval
 in which the pion field is correlated in
isospin, is finite, as one might expect,
and could be as large as 2 to 3. The time
evolution of a non-expanding quenched system
 in $1+3$ dimensions has been studied
numerically in refs. 4 and 10. The authors
observe indeed a rather dramatic amplification
 of long wavelength pion modes in the period
immediately following the quench. Analogous
numerical simulations, but taking into account
 expansion, are being done in Cracow $^{11)}$.
\par
The authors of refs. 10, 12 have pointed
out that the correlation length extracted
 from the correlator $\langle \pi(\vec{x},t)
 \pi(\vec{0},t) \rangle$ is not large,
typically of the order of 1 to 2 fm. They
 conclude that large domains of DCC are
unlikely to be created following a quench.
 Consequently the yield of pions from DCC
decay is small and it is very difficult
 to distinguish DCC formation from a trivial
 statistical fluctuation. A way out has been
 suggested in ref. 13, where an "annealed"
scenario has been suggested. Actually, all
these authors are {\em averaging over the initial conditions}. But we have
argued before that
 this may be misleading. It is plausible
that generically the bubble is hardly
observable. The truly relevant question
is: what is the {\em probability} that
a large enough domain is formed, when
one starts with a random initial state?
Provided this probability is not too
small, the signal of DCC formation may
 be above the trivial backgroud. In principle, the probability in question, an
analogue of
(\ref{prob}),  can be found from numerical
 simulations while the background can be
 estimated using standard models of
multiparticle production. Such a useful
 phenomenological analysis has not been
carried out yet.  Notice, that (\ref{law})
 is expected to hold within the sample of
 (perhaps) rare events corresponding to
the formation of large domains of DCC,
since the isospin orientation of the
condensate is not expected to be correlated
 with its spacial extent.
\par
One should mention, that an experiment
$^{14)}$ is presently been run at the
TEVATRON, with the aim of observing a
disoriented chiral condensate. One expects
 that some results will become available
 during this year. Hopefully,  focusing
on very high multiplicity events in $p
\bar{p}$ (instead of heavy-ion) collisions
 one meets the regime required for the
formation of DCC, but this is not granted.
\par
Although DCC formation has not yet been
 established experimentally, this hypothetical
 phenomenon has already triggered an intense
theoretical activity, being a good pretext
for studying non-equilibrium aspects of
complicated high-energy nuclear collisions.
 For a long time, following the seminal work of Landau~$^{15)}$ on hadron
hydrodynamics,
 the subject has been dominated by the
concept of (local) thermal equilibrium.
The increasing attention devoted to the
 non-equilibrium aspects of the problem
 is certainly a very interesting developpment.

\pagebreak
\par\noindent
{\bf References}
\baselineskip 0.45cm
\begin{enumerate}
\item A.A. Anselm, Phys. Lett. B217,
 169 (1989); A.A. Anselm and M. Ryskin,
 Phys. Lett. B266, 482 (1991)
\item J.D. Bjorken, Int. J. Mod. Phys.
 A7, 4819 (1992); Acta Phys. Polon.
 B23, 561 (1992).
\item J.P. Blaizot and A. Krzywicki,
 Phys. Rev. D46, 246 (1992).
\item K. Rajagopal and F. Wilczek,
 Nucl. Phys. B399, 395 (1993); Nucl.
Phys. B404, 577 (1993).
\item J.P. Blaizot and A. Krzywicki,
Phys. Rev. D, to be
published (February 1994, hep-ph/9402274).
\item K.L. Kowalski and C.C. Taylor,
  CWRUTH-92-6    (June 1992), hep-ph/9211282.
\item  See C.M.G. Lattes, Y. Fujimoto
 and S. Hasegawa, Phys. Rep. 65, 151 (1980).
\item W. Heisenberg, Z. Phys. 133, 65 (1952).
\item Z. Huang and X.-N. Wang, LBL
preprint, LBL-34931 ( December 1993).
\item S. Gavin, A. Gocksch and R.
Pisarski, Brookhaven preprint, BNL-GGP-1
(October 1993).
\item A. Bia\l as, private communication.
\item D. Boyanovsky, H.J. de Vega and R.
Holman, Pittsburgh preprint PITT-94-1
(January 1994).
\item S. Gavin and B. M\"{u}ller,
Brookhaven preprint, BNL-GM-1 (December 1993).
\item See J.D. Bjorken, Stanford
preprint SLAC-PUB-6430 (March 1994).
\item L.D. Landau, Izv. Akad. Nauk
SSSR, Ser. Fiz. 17, 51 (1953).
\end{enumerate}
\end{document}